# Programmable Interface for Statistical & Simulation Models (PRISM): Towards Greater Accessibility of Clinical and Healthcare Decision Models

Amin Adibi, MSc[1]; Stephanie Harvard, PhD[1]; Mohsen Sadatsafavi, MD, PhD[1]

Affiliations:

[1] Respiratory Evaluation Sciences Program, Faculty of Pharmaceutical Sciences, University of British Columbia, 2405 Wesbrook Mall, Vancouver, British Columbia, Canada V6T 1Z3.

**Corresponding author:** Mohsen Sadatsafavi
Associate Professor, Faculty of Pharmaceutical Sciences
Centre for Heart Lung Innovation & Dept of Medicine (Respirology)
University of British Columbia
Email: msafavi@mail.ubc.ca

This study was funded by a John R. Evans Leaders Fund from the Canadian Foundation for innovation, and an arm's length research contract from the BC Academic Health Sciences Network. The funding agreements ensured the authors' independence in designing the study, interpreting the data, writing, and publishing the report.

**Abstract**

**Background:** Increasingly, decision-making in healthcare relies on computer models, be it clinical prediction models at point of care or decision-analytic models at the policymaking level. Given the important role models play in both contexts, their structure and implementation be rigorously scrutinized. The ability to interrogate input/output associations without facing barriers can improve quality assurance mechanisms while satisfying privacy/confidentiality concerns and facilitating the integration of models into decision-making. This paper reports on the development of Programmable Interface for Statistical & Simulation Models (PRISM), a cloud-based platform for model accessibility.

**Methods:** PRISM emphasizes two main principles: 1) minimal specifications on the side of model developer to make the model fit for cloud hosting, and 2) making client access completely independent of the resource requirement and software dependencies of the model. The server architecture integrates a RESTful Application Programming Interface (API) infrastructure, JSON for data transfer, a routing layer for access management, container technology for management of computer resources and package dependencies, and the capacity for synchronous or asynchronous model calls.

**Results:** We discuss the architecture, the minimal API standards that enable a universal language for access to such models, the underlying server infrastructure, and the standards used for data transfer. An instance of PRISM is available as a service via the Peer Models Network http://peermodelsnetwork.com. Through a series of case studies, we demonstrate how interrogating models becomes possible in standardized fashion, in a way that is irrespective of the specifics of any model.

**Conclusions:** We have developed a publicly accessible platform and minimalist standards that facilitate model accessibility for both clinical and policy models.

**Keywords:** accessibility, transparency, modeling in healthcare, disease modeling, decision-analytics, health economics

## Background

Medical decision making increasingly relies on computational models. In clinical practice, clinical prediction models assist physicians in diagnosing patients and planning the course of treatment based on the projected prognosis of individual patients. Governments and regulators rely on decision-analytic or health economics models to decide which competing heath technologies provide the best outcome for the money spent, while public health units use infectious disease and environmental exposure models to plan and guide exposure control strategies.

Models typically represent substantial intellectual and monetary investments. They are often prototyped by academic, government, or industry researchers with the intention of producing the results that are published in academic papers or reports, integrated into routine care, submitted to approval agencies, or used to inform specific technology adoption or policy decisions within a jurisdiction. Developing well-documented, optimized, and easy-to-use software is rarely the goal of modeling units in such organizations, and hardly within their means.

In this manuscript, we focus on a particular challenge facing all these modelling efforts: difficulties in in accessing model outputs. As we describe, this lack of model *accessibility* prevents thorough model interrogation, resulting in lack of trust and potentially inaccurate predictions, causes unnecessary repetition leading to wasteful research, and slows the uptake and application of such models. We briefly review efforts to overcome these challenges as they apply to modelling in healthcare, and propose PRISM, a cloud-based model accessibility platform for models developed in R.

### Model Accessibility as a form of Transparency

Clinical prediction and policy models should be scrutinized before their output informs decisions that can potentially affect millions of lives. To support this process, reporting guidelines for clinical prediction models [1] and health economic evaluations [2] recommend providing details on the rationale and basis for model assumptions and the types of validation undertaken, and sufficiently transparent reporting to enable others to replicate the model, evaluate its validity, and assess the model's overall usefulness to inform decision-making. Others have called for making models open-source [3] and following standard coding practices [4] to improve transparency and facilitate model-sharing. A particularly laudable effort is the Global Health Cost-Effectiveness Analysis Registry that provides a growing database of open-source models (http://ghcearegistry.com/orchard/about-the-clearinghouse) [5].

One possibility is to move beyond reporting and mere code-sharing and facilitate direct access to models, providing both technical and non-technical stakeholders the opportunity to interrogate models and explore different sets of inputs and outputs and the impact of altering model assumptions. In the context of modeling, 'accessibility', the ability of an end-user to interrogate the input/outputs of a model, is a particularly robust and often overlooked form of 'transparency'.

There are various reasons for pursuing accessibility as a form of transparency, for both clinical predication and decision-analytic models. For one, increasingly, attention is being paid to social and ethical value judgments in modelling [6–8], leading to calls for greater reflection and information on models' adequacy-for-purpose from the perspective of different stakeholders. Even when the models are open source, many stakeholders lack the technical knowledge and means to access and examine the model for themselves. A lack of accessibility also hampers further developments, re-use, and clinical implementation of models, and leads to repetition and wasteful research. We have previously called for a *survival-of-the-fittest* approach towards validation of healthcare models, where multiple clinical prediction models can be set up in a pipeline to receive real-time streaming data (e.g., from an electronic medical records (EMR) system) and compete for better predictive performance over time [9]. An analogous setup for policy models is prospective testing of model projections with the arrival of new studies, as it is practiced by the Mt Hood Diabetes Challenge Network [10]. Direct model access would greatly facilitate not only these initiatives but also integration of healthcare models into other pieces of software and hardware, including but not limited to diagnostic devices, EMR systems, web and smart phone applications, and software packages used for research.

### *Scientific Reproducibility*

Lacking direct model access, peer-reviewers, policymakers, and other users of models often rely on selected results from a model published in a manuscript or report to judge the quality and usefulness of the model. While some clinical prediction models are simple and can easily be programmed by the user, others require complex computations. Many decision models are so complex that it would be nearly impossible to fully document their inner workings (to the point of enabling reproducibility) within the confines of a typical scientific report. While an increasing push for transparency has compelled many modellers to make their models open-source, it is often the case that very few people, other than the developers of the model, are practically able to quality-certify a model based on its code. This is particularly the case for the larger and more complex 'whole disease' models that are being advocated for to reduce duplicate efforts in decision modeling and represent the broader context of decision problems [11]. It is not feasible for a peer-reviewer to vet, in limited time, the internal and external validity of a large decision model that was developed over the course of several years by several developers. Even when the code is made available, a common challenge is in getting a complex model to run on the local computing environment of the user, due to the potential need for specialized software environment or hardware specification.

### *Integration of models into routine care*

For clinical prediction models, the need for accessibility stems from the direct applicability of such models at point of care, such as integration within Clinical Decision Support Systems and EMR. Currently, clinicians might find themselves limited to models provided by a certain commercial entity. A troubling trend in recent years has been an increasing number of under-validated models or proprietary models embedded in EMR systems that have not gone through extensive peer review and independent validation [12]. If all clinical prediction models in the same clinical area could be accessed in standardized ways, then health providers could make an informed decision to pick the model, if any, that would have the best performance in their local setting, while being able to monitor troublesome trends such as dataset shift and calibration drift in real-time [13,14].

### *Confidentiality and intellectual property*

Several challenges need to be addressed before enabling public access to models and some of these challenges are common to both clinical prediction and policy models. Some models might require on-demand access to individual patient data that cannot be made public. For example, a clinical prediction model for lung function decline handles missing predictors by fitting a new model on the original development data 'on demand' [15]. It is also common for cost-effectiveness models, notably those submitted to health technology assessment agencies by pharmaceutical companies, to contain intellectual property and proprietary information that the sponsor is unwilling to share with the public. A practical model accessibility platform should be able to protect confidential information such as patient data and confidential pricing.

## Some available solutions

Health technology assessment models are traditionally implemented in Microsoft Excel, which is not an ideal platform for complex model development [16]. While models built in Excel can be shared, they are incredibly difficult to test and debug, offer little in the way of version control and documentation of model development, and at times require the user to have high-end processing capabilities or face very long running times. Occasionally, web applications or dashboard prototypes that accompany the submission help end-users interactively explore some of the results. Well-resourced modeling efforts might have their customized implementation solutions. An example is the OncoSim models in Canada [17].

Clinical prediction models that are picked up by guidelines such as the Atherosclerotic Cardiovascular Disease Risk Estimator[18] usually have publicly accessible web portals. While web apps do make healthcare models more accessible, professional development and deployment of web apps is often too expensive for model development teams and in-house-developed web apps often come with poorly designed user interfaces and unreliable back-ends that affect user experience. Further, even professionally developed web applications still do not provide programmatic access to the model required for high-throughput validation and testing.

Recently, the Shiny web app framework for R programmers has attained substantial popularity among decision modelers [19]. Shiny provides customized web interface to R functions and can enable rich, event-driven interactive web application and is more accessible to researchers compared with common web programming languages. However, this platform does not offer programmatic access through standard Application Programming Interfaces (APIs) access to models (unless the user signs up for the costly premium RStudio Connect service) and as such, does not enable large-scale model validation and implementation. There are some other commercial entities that offer programmable APIs, for example, the *evidencio* (https://www.evidencio.com/home) platform that provides standard web portals and API for some clinical prediction models, but any professional use of this service requires payment.

### Our proposed approach: Separation of concerns for model developers and model users

We believe model developers and users should be able to focus on the aspects of modeling that concern them, not on the specifics of model deployment or access. On the side of model developers, who are often experts in the science of modeling but are not software engineers or programming experts, this means that no additional requirement for making a model available is required. On the side of the end-users, they can now focus on interrogating or using a model, rather than nuances of how each model can be accessed. Such separation of concerns is achieved by providing an accessibility platform that imposes minimal requirements for the developers and a unified interface for clients that hides the technical complexity and dependencies of particular models.

## Programmable Interface for Statistical & Simulation Models (PRISM)

We have taken an early attempt on this front by creating the Programmable Interface for Statistical & Simulation Models (PRISM). PRISM is a readily available cloud-based web API service that puts R models on the cloud and provide access to them through standardized http calls. As a motivating example, consider the following curl command and its response:

```
curl \
-X POST \
-H "x-prism-auth-user: REPLACE_WITH_API_KEY" \
-H "Content-Type: application/json" \
-d
'{"func":["prism_model_run"],"model_input":[{"male":1,"age":70,"smoker":1,"FEV1":2.5,"height":1.68,
"weight":65}]}' \
https://prism.peermodelsnetwork.com/route/fev1/run
```

["{\"Year\":[0,1,2,3,4,5,6,7,8,9,10,11,12,13,14,15,0,1,2,3,4,5,6,7,8,9,10,11,12,13,14,15],\"FEV1\":
[2.5,2.3865,2.3192,2.251,2.1819,2.1119,2.0411,1.9694,1.8968,1.8233,1.7489,1.6737,1.5976,1.5206,1.44
27,1.3639,2.5,2.5167,2.4749,2.4322,2.3886,2.3442,2.2988,2.2526,2.2056,2.1576,2.1088,2.059,2.0084,1.
957,1.9046,1.8513],\"variance\":[0,0.0156,0.0181,0.0222,0.0278,0.0349,0.0435,0.0537,0.0654,0.0786,0
.0933,0.1096,0.1274,0.1467,0.1676,0.1899,0,0.0156,0.0181,0.0222,0.0278,0.0349,0.0435,0.0537,0.0654,
0.0786,0.0933,0.1096,0.1274,0.1467,0.1676,0.1899],\"FEV1_lower\":[2.5,2.1418,2.0553,1.959,1.8552,1.
7458,1.6321,1.5152,1.3956,1.2738,1.1501,1.0248,0.898,0.7698,0.6404,0.5097,2.5,2.272,2.211,2.1402,2.
0619,1.978,1.8899,1.7985,1.7044,1.6081,1.5099,1.4101,1.3088,1.2062,1.1023,0.9972],\"FEV1_upper\":[2
.5,2.6312,2.583,2.5429,2.5086,2.4781,2.45,2.4236,2.398,2.3728,2.3477,2.3226,2.2971,2.2713,2.245,2.2
181,2.5,2.7613,2.7387,2.7241,2.7153,2.7103,2.7078,2.7068,2.7068,2.7071,2.7076,2.7079,2.708,2.7077,2
.7069,2.7055],\"Scenario\":[\"Smoking\",\"Smoking\",\"Smoking\",\"Smoking\",\"Smoking\",\"Smoking\"
,\"Smoking\",\"Smoking\",\"Smoking\",\"Smoking\",\"Smoking\",\"Smoking\",\"Smoking\",\"Smoking\",\"
Smoking\",\"Smoking\",\"QuitsSmoking\",\"QuitsSmoking\",\"QuitsSmoking\",\"QuitsSmoking\",\"QuitsSm
oking\",\"QuitsSmoking\",\"QuitsSmoking\",\"QuitsSmoking\",\"QuitsSmoking\",\"QuitsSmoking\",\"Quit
sSmoking\",\"QuitsSmoking\",\"QuitsSmoking\",\"QuitsSmoking\",\"QuitsSmoking\",\"QuitsSmoking\"]}"]

The above command calls a clinical prediction model of lung function in COPD patients [20]. The server returns the results including projected lung function over the next 15 years under the two scenarios of continuing to smoke or quitting smoking. The user gets such results without having to install R, the R package that implements the model (and its long list of dependencies) and does not have to be worried about computational resources on their computer. Given such *http* calls are now implemented in standard libraries in many programming environments (e.g., the *httpr* package for R, or *request* library for python), access to a model like this can be done in any programming environment and the results be processed (e.g., turned into a graph) as the user demands.

**Principles**

The major design principles underlying PRISM include:

1. **Web-API access**. Models that sit on the cloud are easily updated by developers and accessed by end-users. Testing, debugging, and package dependencies are all managed on the cloud and behind the scene.

2. **Minimal specification**. This principle recognizes that model developers should worry about the content of their model and not about how easily it can be accessed on the cloud. Accordingly, PRISM in its core will only require the model to be turned into a standard R package, with a minimal set of functions exposed in a standardized way. Creating comprehensive standards for model accessibility requires coordinated efforts by the broader community.

3. **Scalability**: the proposed solution should be ready to host models of any level of complexity. As well, conflicts due to the dependency of the models on additional components (e.g., specific packages) should be avoided by design.

4. **Access managemen**t: while generally models should be open-source and publicly accessible, the platform should respect the potential needs of model developer including the protection of intellectual property or hiding private data.

The above design principles distinguish PRISM from other means of facilitating model accessibility, reviewed above, and such differences are summarized in Table 1.

*Table 1 Comparing features of the PRISM platform with alternatives*

| **Functionality** | **PRISM** | **Code sharing** | **Shiny Server Open-Source** | **Shinyapps.io** | **RStudio Connect** |
| --- | --- | --- | --- | --- | --- |
| Interactive Access to Results | Yes | Yes | Yes | Yes | Yes |
| Secure Programmatic Access | Yes | Possible but expensive | No | No | Yes |
| Standard Input/Output Structure | Yes | No | No | No | No |
| Authentication, Levels of Access, and Logging | Yes | No | No | Yes | Yes |
| Synchronous and Asynchronous Calls | Yes | No | No | No | Yes |
| Container Management | Yes | No | No | No | No |
| Accessible Interface in R | Yes | No | No | No | Yes |
| Accessible Interface in Excel | Yes | No | No | No | No |
| Cost | Free | Free | Free | Expensive | Very Expensive |

## Architecture

***Figure 1*** provides a schematic illustration of the proposed technology. Individual components are discussed below.

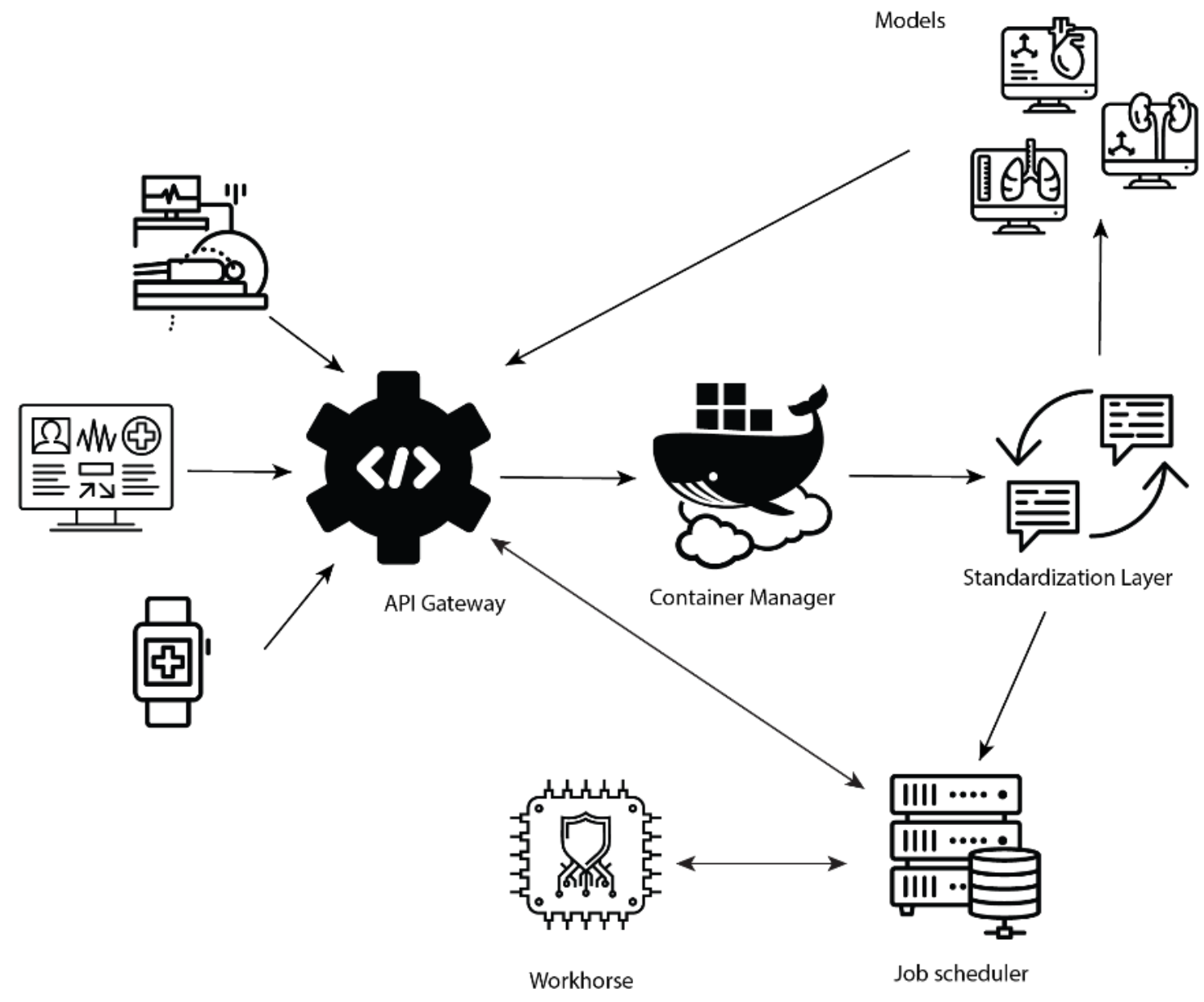

*Figure 1 Architecture of the PRISM*

**Core technologies: Open CPU, JSON, Access management, Containers, and Asynchronous access**

**1. OpenCPU:** The core server technology is OpenCPU, an open-source platform enabling remote procedure calls to R[21]. Open CPU all exposed functions of a given R package (i.e., the model) through HTTP calls. Function calls are typically enabled via synchronous HTTP POST requests. The request will return the result of the R function call as well as information about the call in a temporary web address. OpenCPU also automatically captures the graphical output of the R function call and makes it available in the temporary address. OpenCPU provides parallel processing as well as sandboxing (preventing unauthorized access to key system resources by unauthorized code). Open-CPU is by default a stateless synchronous platform, indicating that upon receiving a call request, it will start an R process, process the request, return the results, and terminate the R process. This contrasts with state-based platforms like Shiny that keep the R session open for as long as the user is connected.

**2. JSON:** Data are transferred between the client and server in JSON (Javascript Object Notation) format. JSON is an industry-standard data coding format that is implemented in many programming languages and data environments, including in R (e.g., jsonlite package)[22].

**3. Access layer:** OpenCPU by default exposes all functions in a package and makes their source code visible. While this is advantageous for many publicly available models, it cannot *per se* accommodate privacy and confidentiality requirements. Further, OpenCPU does not provide, by default, any quota on the use of system's resources by a given user (so submitting several demanding requests can effectively shut down the server). The access layer, implemented in the freely available CapRover software [23], enables user management and logging of information. User management is performed through API keys. Each API key is unique to each user and enables different levels of access to the model. The API key also enables monitoring and logging of model calls (for example, specifying a quota on CPU usage such that a particular user does not overwhelm the server). API keys are passed to the server via standard http headers.

**4. Docker Containers:** Each model, encapsulated in an R package, can have unique dependencies. Dependencies of models can conflict with each other; updating an R dependency for one package can break another package. And generally installing all R packages in a single R environment will generate difficulties in resource management for both the human administrators and the computing environment. Container technology enables different instances of applications (in this case an R environment and its dependencies) access operating system resources but be kept separate from each other. PRISM enables a flexible structure, ranging from several models installed in the same R environment to each model having its own R environment.

A Docker management system controls container apps, while an API router system takes care of user authentication, security, and logging. The platform can support both open-source and proprietary models with various levels of access for different users. A standardization layer provides wrapper functions that enable calling all models with a single standard API syntax.

**5. Asynchronous model call**: OpenCPU by default provides synchronous communication: the http request on the client side will have to wait until this process is completed. For computationally fast models, this generally takes a few seconds. For computationally intensive models, however, this approach will not be practical. To accommodate this, PRISM hosts a 'work horse' architecture for computationally heavy models. For such models, the user can send a request as an asynchronous job (via relevant parameters in the http call header). If the particular model has an asynchronous implementation, the request is then put on a queue and a ticket is returned to the client. The ticket can be used later to check the progress of the job and retrieve results once the computation is concluded. Alternatively, the http call can specify an email address which will be used to notify the user once the computations are completed. As an example, consider the following curl command:

```
curl \
-X POST \
-H "x-prism-auth-user: REPLACE_WITH_API_KEY" \
-H "Content-Type: application/json" \
-d \
'{"func":["prism_model_run"],"email_address":["REPLACE_WITH_YOUR_EMAIL"]}'\
https://prism.peermodelsnetwork.com/route/epic/async/run
```

```
Response:
["{\"token\":[\"b3nd7kbuk5\"],\"email_address\":[\"YOUR_EMAIL\"],\"error_code\":[0]}"]
```

An error code of 0 indicates that the request was submitted successfully. Note that the call address indicates the asynchronous nature of this request.

## Implementation

We deployed an instance of the PRISM server as a part of the Peer Model Network website, available online at PeerModelsNetwork.com. A list of models currently hosted on the platform can be seen at models.peermodelsnetwork.com. While this implementation of the cloud-based PRISM software infrastructure is hosted at the University of British Columbia, the platform can be deployed on all commonly used cloud systems including Microsoft Azure, Amazon Web Services, Google Cloud, Digital Ocean, as well as on-premises Linux servers. This enables regulatory agencies that have strict security and privacy requirements to deploy and use the platform internally, if need be.

**Current Standardization**

One of the challenges of working with many models is that each model requires the user to follow a different syntax and workflow in terms of the functions and the specific order in which they need to be called, and the input variables and parameters that need to be provided. The PRISM solution to this problem is a standardization wrapper layer for each model, which enables the user accessing the cloud to follow a standard workflow with a unified syntax when calling different models on the cloud server.

Our guiding principle in designing minimal standardization requirement was such that no more standards are imposed on the modeler.

Before a model developed in R can be hosted on PRISM, the model and the wrapper file should be turned into a standard R package if it is not already in that format [24]. For an example model package, see the source code of the R package for the Acute COPD Exacerbation Prediction Tool at https://github.com/resplab/accept [25]. There are several tools for turning a set of R functions to an R package without the end-user requiring a deep understanding of the internal structure of R packages [24]. For models that have previously been published as an R package, the same package can be used for PRISM, as a generic secondary R package (bridgePrism, https://github.com/resplab/bridgePrism) can interact with the R model.

## Core API Functions

For the model to be accessilbe, at the minimum, the package should expose one function, *prism_model_run(model_input)*. This function must take model_input as a list, call the model with the appropriate model-specific command and return results as a list.

The function *prism_get_default_input()* should return the 'default' input set of the model. For decision-analytic models, this generally represents the base case analysis. For clinical prediction models, the input can represent the profile of a typical person to whom the prediction model applies. The main purpose of this function is to facilitate the process of learning the input structure that model for the user. The user can then change this input and call the model with modified inputs. Table 2 summarizes the standard model call functions defined in the PRISM wrapper package.

*Table 2 Summary of standardized model call functions*

| Standard Command | Details |
|---|---|
| *prism_get_default_input* | Returns default input structure and values for the specific model. For decision-analytic models, this generally represents the base case analysis. For clinical prediction models, the input can represent the profile of a typical person to whom the prediction model applies. The main purpose of this function is to facilitate the process of learning the input structure that model needs to make prediction. |
| *prism_model_run* | Runs the model with the given input parameter and returns the results, either synchronously or asynchronously |
| *prism_get_async_results* | Checks the status of a previously submitted asynchronous job (for models with asynchronous call capability) |

In general, the standard workflow for programmatic access to models on the *peermodels* cloud includes the following steps:

1. Retrieving default input structure and values.
2. Modifying default input values (if needed) and submitting the inputs to get model results.

The first step is designed to familiarize the user to the input parameters and their default values in the model of choice. This step simplifies this process as different types of models might have anywhere between a handful (as in simple clinical calculators) to many thousands of nested input parameters (as in complex whole-disease policy models). If the user is already familiar with the input structure of the model, they can skip the first step and directly call the model with the input value provided.

Depending on how computationally intensive a model might be, the model will return results instantly (i.e. synchronously) or with a delay (asynchronously). The architecture of these two modes of running models is described below.

### Exemplary scenarios for client access

### R Package *peermodels*

The R package `peermodels` provides wrapper functions that simplify access to models on the Peer Models Network cloud. The `peermodels` package is particularly useful for data wrangling, processing customized web apps, and statistical analysis of the outputs of a model. The case studies in the later sections of this manuscript provide examples of such scenarios. The `peermodels` package is available on the Comprehensive R Archive Network (CRAN)[26].

### Case studies

### Decision Model Case Study: Evaluation Platform in COPD (EPIC)

The Evaluation Platform in COPD is a Canadian decision-analytic model that uses discrete-event simulation modelling to model different pathways of care in COPD in order to evaluate policy choices by projecting health and cost outcomes at the societal level [27]. EPIC is currently deployed on the PRISM server and is available for both synchronous and asynchronous calls. EPIC is accessible through the `peermodels` package in R, as well as an Excel spreadsheet and standard RESTful API calls from other programming languages. The following code snippet in R retrieves default inputs of this model:

```
> library(peermodels)
> input <- get_default_input(model_name = "epic", api_key = "YOUR_API_KEY")
Calling server at https://prism.peermodelsnetwork.com/route/epic/run
> length(input)
[1] 105

> head(input)
$global_parameters.age0
[1] 40

$global_parameters.time_horizon
[1] 20

$global_parameters.discount_cost
[1] 0.03

$global_parameters.discount_qaly
[1] 0.03

$agent.p_female
[1] 0.5

$agent.height_0_betas
       [,1]    [,2]    [,3]     [,4]   [,5]
[1,] 1.8266 -0.1309 -0.0012 2.31e-06 -2e-04
```

In this model, the default input is a list of 103 parameters, some of which are a matrix themselves. Using the acquired default input or a modified version of it, the model can be queried either synchronously or asynchronously. For synchronous calls:

```
> results <- model_run(model_input = input, model_name = "epic", api_key = "YOUR_API_KEY")
Calling server at https://prism.peermodelsnetwork.com/route/epic/run
> names(results)
 [1] "status"                          "n_agents"
 [3] "cumul_time"                      "n_deaths"
 [5] "n_COPD"                          "total_exac"
 [7] "total_exac_time"                 "total_pack_years"
 [9] "total_doctor_visit"              "total_cost"
[11] "total_qaly"                      "total_diagnosed_time"
[13] "n_alive_by_ctime_sex"            "n_alive_by_ctime_age"
[15] "n_smoking_status_by_ctime"       "n_current_smoker_by_ctime_sex"
[17] "sum_fev1_ltime"                  "cumul_time_by_smoking_status"
[19] "cumul_non_COPD_time"             "sum_p_COPD_by_ctime_sex"
[21] "sum_pack_years_by_ctime_sex"     "sum_age_by_ctime_sex"
[23] "n_death_by_age_sex"              "n_alive_by_age_sex"
[25] "sum_time_by_ctime_sex"           "sum_time_by_age_sex"
[27] "sum_weight_by_ctime_sex"         "n_COPD_by_ctime_sex"
[29] "n_COPD_by_ctime_age"             "n_inc_COPD_by_ctime_age"
[31] "n_COPD_by_ctime_severity"        "n_COPD_by_age_sex"
[33] "n_Diagnosed_by_ctime_sex"        "n_Overdiagnosed_by_ctime_sex"
[35] "n_Diagnosed_by_ctime_severity"   "n_total_case_detection"
```

```
[37] "cumul_time_by_ctime_GOLD"              "n_exac_by_ctime_age"
[39] "n_severep_exac_by_ctime_age"           "n_exac_death_by_ctime_age"
[41] "n_exac_death_by_ctime_severity"        "n_exac_death_by_age_sex"
[43] "n_exac_by_ctime_severity"              "n_exac_by_gold_severity"
[45] "n_exac_by_gold_severity_diagnosed"     "n_exac_by_ctime_severity_female"
[47] "n_exac_by_ctime_GOLD"                  "n_exac_by_ctime_severity_undiagnosed"
[49] "n_exac_by_ctime_severity_diagnosed"    "n_GPvisits_by_ctime_sex"
[51] "n_GPvisits_by_ctime_severity"          "n_GPvisits_by_ctime_diagnosis"
[53] "n_cough_by_ctime_severity"             "n_phlegm_by_ctime_severity"
[55] "n_wheeze_by_ctime_severity"            "n_dyspnea_by_ctime_severity"
[57] "n_mi"                                  "n_incident_mi"
[59] "n_mi_by_age_sex"                       "n_mi_by_ctime_sex"
[61] "sum_p_mi_by_ctime_sex"                 "n_stroke"
[63] "n_incident_stroke"                     "n_stroke_by_age_sex"
[65] "n_stroke_by_ctime_sex"                 "n_hf"
[67] "n_incident_hf"                         "n_hf_by_age_sex"
[69] "n_hf_by_ctime_sex"                     "medication_time_by_class"
[71] "n_exac_by_medication_class"

> head(results)
$status
[1] 0

$n_agents
[1] 96428

$cumul_time
[1] 1332577

$n_deaths
[1] 20539

$n_COPD
[1] 14512

$total_exac
[1] 31431  6027 10776   964
```

An asynchronously will generate a token which can be used later to retrieve the results:

```
> model_run(input = input, model_name = "epic", api_key = "YOUR_API_KEY", async = TRUE)
Calling server at https://prism.peermodelsnetwork.com/route/epic/async/run

$token
[1] "s0z00bh4qd"

$error_code
[1] 0
```

The above token can then be called through get_async_results to see if the results are ready.

```
> results_async <- get_async_results(model_name = "epic", token = "s0z00bh4qd")
Calling server at https://prism.peermodelsnetwork.com/route/epic/run
> results_async$status
[1] "[COMPLETED]"

> length(results_async$status_data)
[1] 71

> head(results_async$status_data)
```

```
$status
[1] 0

$n_agents
[1] 96428

$cumul_time
[1] 1331888

$n_deaths
[1] 20650

$n_COPD
[1] 14597

$total_exac
[1] 31860  6073 10679   962
```

**Clinical Model Case Study: Acute COPD Exacerbation Prediction Tool (ACCEPT)**

Acute COPD Exacerbation Prediction Tool (ACCEPT) is a clinical prediction model that projects the rate of all and severe exacerbations of COPD within the next year[25].

The following code snippet in R can get the default input structure for this model.

```
> library(peermodels)
> input <- get_default_input(model_name = "accept", api_key = "YOUR_API_KEY")
Selected model is accept
Calling server at https://prism.peermodelsnetwork.com/route/accept/run
This is acceptPrism - PRISM enabled!
Selected model is accept
Calling server at https://prism.peermodelsnetwork.com/route/accept/run
```

Compared to the previous example, this prediction model has a more manageable input that has 21 variables:

```
> input
     ID male age smoker oxygen statin LAMA LABA ICS FEV1 BMI SGRQ LastYrExacCount
1 10001    1  70      1      1      1    1    1   1   33  25   50               2
  LastYrSevExacCount randomized_azithromycin randomized_statin randomized_LAMA randomized_LABA
1                  1                       0                 0               0               0
  randomized_ICS random_sampling_N calculate_CIs
1              0               100         FALSE
```

The results can then be obtained with a straightforward synchronous call to the model, showing that this particular COPD patient has a predicted severe exacerbation probability of about 50% for the next year:

```
> results <- model_run(input = input, model_name = "accept", api_key = "YOUR_API_KEY")
Selected model is accept
Calling server at https://prism.peermodelsnetwork.com/route/accept/run
> results$predicted_severe_exac_probability
[1] 0.5025
```

## Discussion

The fields of predictive analytics and decision analysis in healthcare are currently at a pivot point. On one hand, clinical prediction is gaining significant momentum under the purview of the so-called "precision medicine" initiatives [28]. Integrating prediction models with increasingly available patient-level data, be it genetic, -omics, administrative, or through EMR systems and smart devices, is bringing about countless possibilities. On the other hand, the recent increase in the use of models in healthcare decision-making has not been without its challenges [29]. Notably, there has been concerns about scientific rigor and validity of models[30], potential conflicts of interests on the part of model developers[31], wasteful repeated attempts to build similar models[9], ethics of modelling and management of value judgments[6,7], and lack of transparency[32].

Efforts to bolster trust in healthcare models have thus far mainly focused on transparency of the modelling process and the final model, improved reporting, and patient and stakeholder involvement. The *peermodels* cloud system presented here focuses on making models more accessible, making it easier to examine the model's assumptions and evaluate its adequacy for purpose. Moreover, programmatic access to models can standardize model validation and integration efforts and make them more efficient. To showcase how this is done in practice, we provide two case studies here, one with a decision-analytic health economics model and another with a clinical prediction model.

We kept the specifications for PRISM at a minimum (with the only mandatory function being prism_model_run). This is a manifestation of our 'separation of concerns' principle and recognizes the general desire among modelers to focus their efforts on proper modeling rather than following sophisticated publishing standards. However, we recognize that standardizing model calls can have substantial benefits. For example, a cost-effectiveness model can follow standards on reporting payoffs, net benefits, as well as common standards for setting the time horizon, running a probabilistic analysis at given times, and so on. If these standards are implemented, common libraries can be developed to generate standard outputs form such models. For example, Value of Information (VoI) analyses that work on standard outputs of a probabilistic analysis can then be called, regardless of the specifics of any model. For clinical prediction models, naming conventions such as those specified by Fast Healthcare Interoperability Resources (FHIR)[33] will facilitate automatic integration of such models into electronic health records, or facilitate automatic external validation, or even recalibration, of such models with the arrival of new data. However, we think specifying such common standards is not a mandate of a model accessibility platform like PRISM and will require broader conversation among all stakeholders in the field. A platform like PRISM can provide a basis for implementation of such standards.

The stateless server technology implemented in PRISM is in contrast with naturally state-based systems such as Shiny. A Shiny app keeps an R process on the server up and running for as long as the user is connected to the application. The benefit of such a solution is the interactive session. For example, the user can run probabilistic analysis and then use the result in multiple instances such as drawing the [cost-effectiveness?] acceptability curve or performing VoI. The downside of this approach is the scalability of the server if multiple users are connected. In PRISM, the emphasis is on secure stateless server calls after which the R process is terminated. A main advantage of this is the as-required consumption of server resources, as well as the R process starting afresh with a new call thus not being at the risk of becoming corrupted with previous function calls.

## Conclusions

Concerns about scientific validity, reproducibility, and management of value judgments in healthcare modelling has led to initiatives that encourage open-source modelling and adherence to reporting guidelines. PRISM cloud infrastructure complements these efforts by providing a standard framework for direct access to healthcare models on the cloud. The PRISM platform enables a wide variety of audiences with different levels of technical expertise to examine models, while at the same time facilitating reproducibility, validation studies, and implementation studies.

References:

1 Moons KGM, Altman DG, Reitsma JB, *et al.* Transparent Reporting of a multivariable prediction model for Individual Prognosis or Diagnosis (TRIPOD): explanation and elaboration. *Ann Intern Med* 2015;**162**:W1-73. doi:10.7326/M14-0698

2 Husereau D, Drummond M, Augustovski F, *et al.* Consolidated Health Economic Evaluation Reporting Standards (CHEERS) 2022 Explanation and Elaboration: A Report of the ISPOR CHEERS II Good Practices Task Force. *Value Health J Int Soc Pharmacoeconomics Outcomes Res* 2022;**25**:10–31. doi:10.1016/j.jval.2021.10.008

3 Cohen JT, Neumann PJ, Wong JB. A Call for Open-Source Cost-Effectiveness Analysis. *Ann Intern Med* 2017;**167**:432–3. doi:10.7326/M17-1153

4 Alarid-Escudero F, Krijkamp EM, Pechlivanoglou P, *et al.* A Need for Change! A Coding Framework for Improving Transparency in Decision Modeling. *PharmacoEconomics* 2019;**37**:1329–39. doi:10.1007/s40273-019-00837-x

5 Center for Evaluation of Value and Risk in Health (CEVR). Global Health Cost Effectiveness Analysis Registry. http://ghcearegistry.org/ghcearegistry/ (accessed 16 Feb 2022).

6 Harvard S, Werker GR, Silva DS. Social, ethical, and other value judgments in health economics modelling. *Soc Sci Med 1982* 2020;**253**:112975. doi:10.1016/j.socscimed.2020.112975

7 Harvard S, Winsberg E, Symons J, *et al.* Value judgments in a COVID-19 vaccination model: A case study in the need for public involvement in health-oriented modelling. *Soc Sci Med 1982* 2021;**286**:114323. doi:10.1016/j.socscimed.2021.114323

8 Winsberg E, Harvard S. Purposes and duties in scientific modelling. *J Epidemiol Community Health* 2022;:jech-2021-217666. doi:10.1136/jech-2021-217666

9 Adibi A, Sadatsafavi M, Ioannidis JPA. Validation and Utility Testing of Clinical Prediction Models: Time to Change the Approach. *JAMA* 2020;**324**:235–6. doi:10.1001/jama.2020.1230

10 Tew M, Willis M, Asseburg C, *et al.* Exploring Structural Uncertainty and Impact of Health State Utility Values on Lifetime Outcomes in Diabetes Economic Simulation Models: Findings from the Ninth Mount Hood Diabetes Quality-of-Life Challenge. *Med Decis Mak Int J Soc Med Decis Mak* 2021;:272989X211065479. doi:10.1177/0272989X211065479

11 Tappenden P, Chilcott J, Brennan A, *et al.* Whole Disease Modeling to Inform Resource Allocation Decisions in Cancer: A Methodological Framework. *Value Health* 2012;**15**:1127–36. doi:10.1016/j.jval.2012.07.008

12 Wong A, Otles E, Donnelly JP, *et al.* External Validation of a Widely Implemented Proprietary Sepsis Prediction Model in Hospitalized Patients. *JAMA Intern Med* 2021;**181**:1065–70. doi:10.1001/jamainternmed.2021.2626

13 Finlayson SG, Subbaswamy A, Singh K, *et al.* The Clinician and Dataset Shift in Artificial Intelligence. *N Engl J Med* 2021;**385**:283–6. doi:10.1056/NEJMc2104626

14 Wong A, Cao J, Lyons PG, *et al.* Quantification of Sepsis Model Alerts in 24 US Hospitals Before and During the COVID-19 Pandemic. *JAMA Netw Open* 2021;**4**:e2135286. doi:10.1001/jamanetworkopen.2021.35286

15 Chen W, Sin DD, FitzGerald JM, *et al.* An Individualized Prediction Model for Long-term Lung Function Trajectory and Risk of COPD in the General Population. *Chest* 2020;**157**:547–57. doi:10.1016/j.chest.2019.09.003

16 Incerti D, Thom H, Baio G, *et al.* R You Still Using Excel? The Advantages of Modern Software Tools for Health Technology Assessment. *Value Health* 2019;**22**:575–9. doi:10.1016/j.jval.2019.01.003

17 Gauvreau CL, Fitzgerald NR, Memon S, *et al.* The OncoSim model: development and use for better decision-making in Canadian cancer control. *Curr Oncol* 2017;**24**:401–6. doi:10.3747/co.24.3850

18 Lloyd-Jones DM, Braun LT, Ndumele CE, *et al.* Use of Risk Assessment Tools to Guide Decision-Making in the Primary Prevention of Atherosclerotic Cardiovascular Disease: A Special Report From the American Heart Association and American College of Cardiology. *J Am Coll Cardiol* 2019;**73**:3153–67. doi:10.1016/j.jacc.2018.11.005

19 Smith R, Schneider P. Making health economic models Shiny: A tutorial. *Wellcome Open Res* 2020;**5**:69. doi:10.12688/wellcomeopenres.15807.2

20 Zafari Z, Sin DD, Postma DS, *et al.* Individualized prediction of lung-function decline in chronic obstructive pulmonary disease. *Can Med Assoc J* 2016;:cmaj.151483. doi:10.1503/cmaj.151483

21 Ooms J. The OpenCPU System: Towards a Universal Interface for Scientific Computing through Separation of Concerns. *ArXiv14064806 Cs Stat* Published Online First: 3 June 2014.http://arxiv.org/abs/1406.4806 (accessed 17 Jun 2021).

22 Pezoa F, Reutter JL, Suarez F, *et al.* Foundations of JSON Schema. In: *Proceedings of the 25th International Conference on World Wide Web*. Republic and Canton of Geneva, CHE: : International World Wide Web Conferences Steering Committee 2016. 263–73. doi:10.1145/2872427.2883029

23 Bigdeli, Kasra. *CapRover*. CapRover 2022. https://github.com/caprover/caprover (accessed 2 Feb 2022).

24 Wickham H, Bryan J. *R Packages*. https://r-pkgs.org/ (accessed 9 Feb 2022).

25 Adibi A, Sin DD, Safari A, *et al.* The Acute COPD Exacerbation Prediction Tool (ACCEPT): a modelling study. *Lancet Respir Med* 2020;**8**:1013–21. doi:10.1016/S2213-2600(19)30397-2

26 Adibi A, Sadatsafavi M. *peermodels: Client-Side R API Wrapper for Peer Models Network Model Repository*. 2021. https://CRAN.R-project.org/package=peermodels (accessed 2 Feb 2022).

27 Sadatsafavi M, Ghanbarian S, Adibi A, *et al.* Development and Validation of the Evaluation Platform in COPD (EPIC): A Population-Based Outcomes Model of COPD for Canada. *Med Decis Making* 2019;**39**:152–67. doi:10.1177/0272989X18824098

28 Collins FS, Varmus H. A new initiative on precision medicine. *N Engl J Med* 2015;**372**:793–5. doi:10.1056/NEJMp1500523

29 Purposes and duties in scientific modelling - PubMed. https://pubmed.ncbi.nlm.nih.gov/35027406/ (accessed 9 Feb 2022).

30 Wynants L, Van Calster B, Collins GS, *et al.* Prediction models for diagnosis and prognosis of covid-19 infection: systematic review and critical appraisal. *BMJ* 2020;**369**:m1328. doi:10.1136/bmj.m1328

31 Garattini L, Koleva D, Casadei G. Modeling in pharmacoeconomic studies: funding sources and outcomes. *Int J Technol Assess Health Care* 2010;**26**:330–3. doi:10.1017/S0266462310000322

32 Sampson CJ, Arnold R, Bryan S, *et al.* Transparency in Decision Modelling: What, Why, Who and How? *PharmacoEconomics* 2019;**37**:1355–69. doi:10.1007/s40273-019-00819-z

33 Mandel JC, Kreda DA, Mandl KD, *et al.* SMART on FHIR: a standards-based, interoperable apps platform for electronic health records. *J Am Med Inform Assoc* 2016;**23**:899–908. doi:10.1093/jamia/ocv189